\documentclass{moriond}
\usepackage{xcolor}
\usepackage{amsmath}
\def\Journal#1#2#3#4{{#1} {\bf #2}, #3 (#4)}

\def\PRD{{\em Phys. Rev.} D}

\def\be{\begin{equation}}
\def\ee{\end{equation}}
\def\bea{\begin{eqnarray}}
\def\eea{\end{eqnarray}}

\newcommand{\Photo}{\includegraphics[height=35mm]{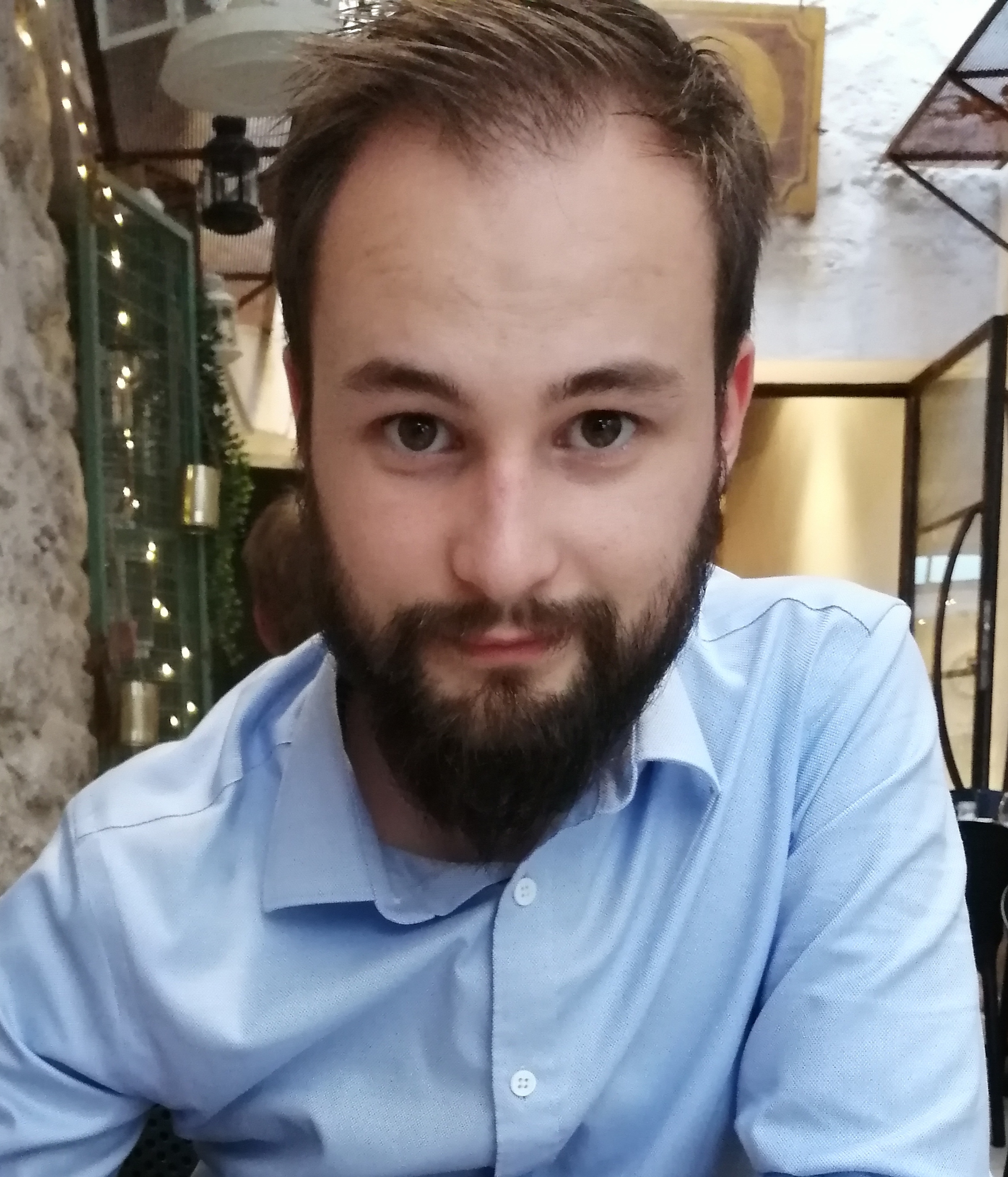}}

\begin{document}
\title{Spectral separation of the stochastic gravitational-wave background for LISA: galactic, cosmological and astrophysical backgrounds}

\author{Guillaume Boileau$^{1}$, Astrid Lamberts$^{1,2}$, Nelson Christensen$^{1}$, Neil J. Cornish$^{3}$, and Renate Meyer$^{4}$}

\address{$^{1}$Artemis, Observatoire de la Côte d'Azur, Université Côte d'Azur, CNRS, 06304 Nice, France\\
$^{2}$Laboratoire Lagrange, Observatoire de la Côte d'Azur, Université Côte d'Azur, CNRS , 06304 Nice, France\\
$^{3}$eXtreme Gravity Institute, Department of Physics,Montana State University, Bozeman, Montana 59717, USA\\
$^{4}$ Department of Statistics, University of Auckland, Auckland, New Zealand}

\maketitle \abstracts{In its observation band, the Laser Interferometer Space Antenna (LISA) will simultaneously observe stochastic gravitational-wave background (SGWB) signals of different origins; orbitally modulated waveforms from galactic white dwarf binaries, a binary black hole produced background, and possibly a cosmologically produced SGWB. We simulate the emission of gravitational waves from galactic white dwarf binaries based on the Lamberts \textit{et al }\cite{Lamberts} distributions and determine a complex waveform from the galactic foreground. We generate the modulated galactic signal detected by LISA due to its orbital motion, and present a data analysis strategy to address it. The Fisher Information and Markov Chain Monte Carlo methods give an estimate of the LISA noise and parameters for the different signal sources. We simultaneously estimate the galactic foreground, the astrophysical and cosmological backgrounds, and estimate detection limits for the future LISA observation of the SGWB in the spectral domain with the 3 LISA channels $ A $, $ E $ and $ T $. In the context of the expected astrophysical background and a galactic foreground, a cosmological background energy density of about $ \Omega_{GW,Cosmo} \approx 8 \times 10^{-13} $ could be detected by LISA with our spectral separation strategy. \cite{Boileau:2021sni}  }

\section{Introduction}

The Laser Interferometer Space Antenna, \textit{LISA}, will simultaneously observe orbital modulated waveforms from galactic white dwarf binaries, a compact binary coalescence produced gravitational-wave background, and potentially a cosmologically created stochastic gravitational-wave background (SGWB). We simulate galactic white dwarf binary gravitational-wave emission based on distributions from various catalogs. We generate the modulated detected galactic white dwarf signal observed by LISA in its orbital motion \cite{Adams}, and present a data analysis strategy to address it. 
The Fisher Information and Markov Chain Monte Carlo methods give an estimation of the \textit{LISA} noise and the parameters for the different signal classes. We simulate a complex waveform from the galactic foreground with $3.5 \times 10^{7}$ binaries. The estimation of the parameters is based on observations in the three LISA channels $A$, $E$ and $T$. We simultaneously estimate the parameters of the noise distribution using a certain LISA noise model. Assuming the presence of an astrophysical background, a cosmological background energy density of around $\Omega_{GW,Cosmo} \approx 1 \times 10^{-12}$ to $1 \times 10^{-13}$ can be detected by LISA. 

\section{Context}

The SGWB is a superposition of a large number of independent sources. In the LISA frequency band [$10^{-5}$, $1$] Hz, we can distinguish different independent sources. The \textbf{galactic foreground}, from double white dwarfs binaries (DWD) in our galaxy~\cite{Lamberts}, will be observed as a modulated waveform (as described with resolved sources to provide the modulation curve for the foreground~\cite{Adams}). The \textbf{astrophysical background} is the SGWB produced by binary black holes (BBH) and binary neutron stars (BNS), and can be predicted from the observations of LIGO/Virgo. For the moment, the amplitude of the astrophysical SGWB is still debated, in fact, the estimation from the LIGO/Virgo observations (O2) give a value~\cite{Chen} of $\Omega_{GW}(25 \text{ Hz}) \simeq 1.8 \times 10^{-9} - 2.5 \times 10^{-9}$, or via simulations from the understanding of BBH and BNS populations~\cite{Perigois}, $\Omega_{GW}(25 \text{ Hz})  \simeq 4.97 \times 10^{-9} - 2.58 \times 10^{-8}$. The last component, and also our main goal, is the SGWB from  \textbf{cosmological sources}, stemming from processes in the early universe, or from cosmic strings. We can summarize the energy spectral density of the SGWB in the LISA band, as a sum of all the SGWBs:  
\begin{equation}\label{eq:SGWB}
  \Omega_{GW}(f) = \frac{A_1 \left(\frac{f}{f_*}\right)^{\alpha_1}}{1 + A_2 \left(\frac{f}{f_*}\right)^{\alpha_2}} + \Omega_{Astro} \left(\frac{f}{f_*}\right)^{\alpha_{Astro}} + \Omega_{Cosmo} \left(\frac{f}{f_*}\right)^{\alpha_{Cosmo}}
\end{equation}
with $\alpha \sim 0$ for the cosmological component, $\alpha \sim 2/3$ for the astrophysical component, and the low-frequency DWD ($\Omega_{DWD,LF}(f) \sim \frac{A_1}{A_2}\left(\frac{f}{f_*}\right)^{\alpha_1-\alpha_2}$). The DWD foreground is assumed to be a broken power law because at high-frequencies ($\simeq 0.1$ Hz) the number of DWDs decreases. 

\section{Simulation and Estimation}
\begin{figure}
\includegraphics[width=150mm]{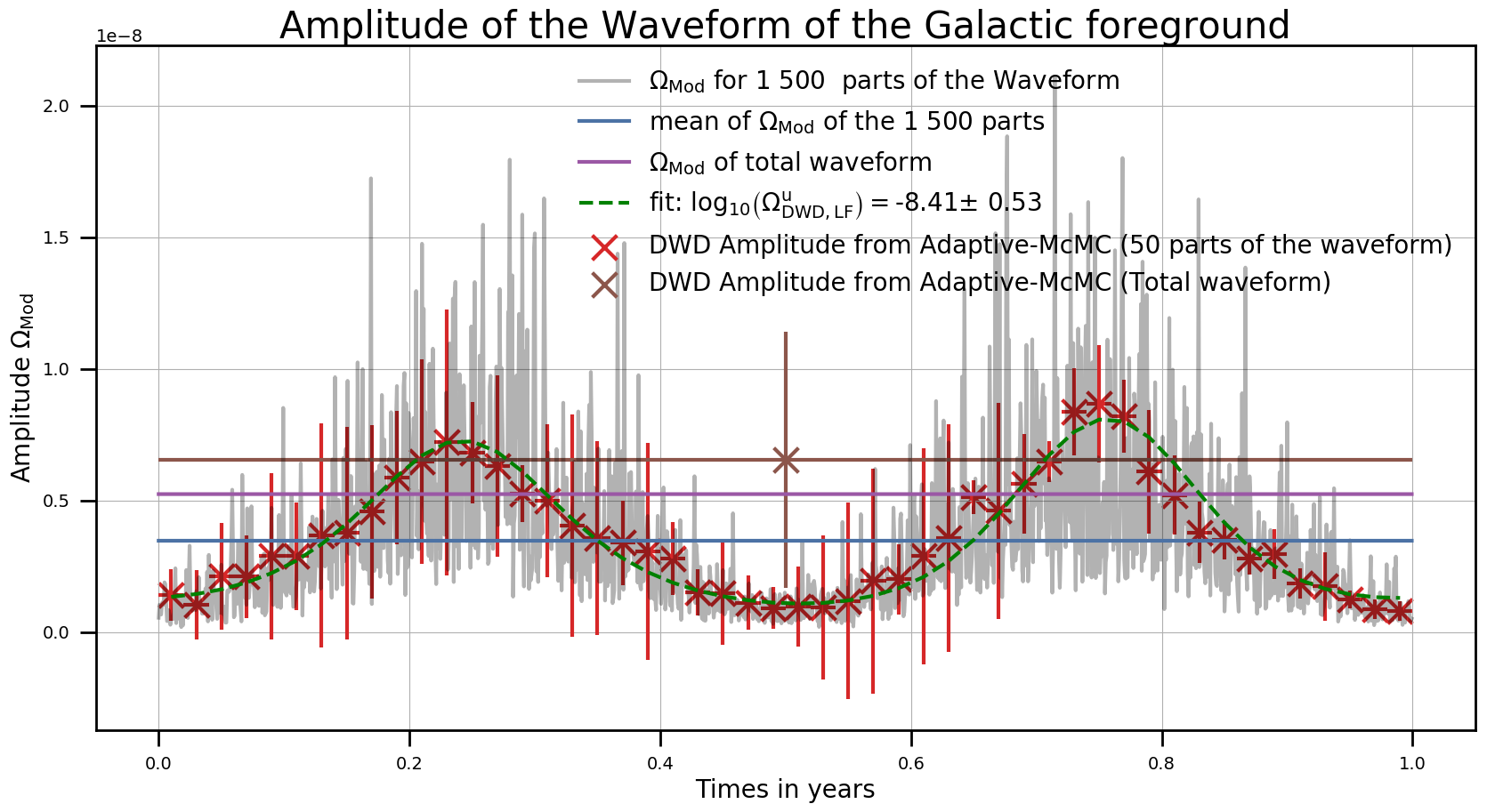}
\caption{Measurement of the orbital modulation of the DWD foreground. In \textbf{\textcolor{gray}{grey}}: 1500 estimates of the amplitude $\Omega_{Mod,i}=  \frac{4\pi^2}{3H_0} \left( \frac{c}{2\pi L}\right)^2 A_i^2$ ($A_i$ amplitude of the characteristic strain). In \textbf{\textcolor{red}{red}}: 50 MCMC of 8 parameters (BBH + WD + LISA noise) for small sections of the year. In \textcolor{green}{green}, fit on the 50 runs to estimate the modulation from the \textit{LISA} antenna pattern 'real' LF DWD amplitude at 3 mHz. Modulation model : $\Omega_{Mod,i} =\Omega_{DWD,LF}^u\left(F^2_{+,i} + F^2_{\times,i} \right)$. }
\label{fig:modulationpro}
\end{figure}

The galactic DWD foreground is generated with a DWD catalog; we use the population from Lamberts {\it et al}~\cite{Lamberts}, and generate the time-series (see Eq.~\ref{eq:waveform}) as a sum of the gravitational waves received by LISA. We assume that the waveform for each binary can be modeled as a pseudo-monochromatic signal:
\begin{equation}\label{eq:waveform}
s(t) = \sum_{i=1}^{N} \sum_{P= +,\times } h_{A,i}(f_{orb,i},M_{1_i},M_{2_i},X_i,Y_i,Z_i,t) \times F_P(\theta,\phi,t) \textbf{D}(\theta,\phi,f)_P:\textbf{e}_{P}
\end{equation}
with $i$ for the binaries; the masses of the two stars being $M_{1_i}$ for the biggest object and $M_{2_i}$ for the smaller; the orbital frequency of the binary is $f_{orb_i}$; the Cartesian position in the Galaxy $X_i,Y_i,Z_i$ and the position in the sky $\theta,\phi$; $F_A$ is the beam pattern function for the polarizations $A = +,\times$, $\textbf{h}_{A,i} = h_{A,i} \textbf{e}_{A}$ the tensor of the amplitude of the gravitational wave; $\textbf{D}$ the one-arm detector tensor; and $\textbf{h}_{A,i}$ the dimensionless GW amplitude. 
The sky seen by the LISA constellation changes during a year; at certain times LISA observes well the center of the galaxy, and sometimes not as well. We need to calculate the modulation of the amplitude of the energy spectral density of the Galactic foreground over time (Eq.~\ref{eq:modulation} and Fig.~\ref{fig:modulationpro}):
\begin{equation} \label{eq:modulation}
    \Omega_{Mod}(t) =\Omega_{DWD,LF}^u\left(F^2_+(t) + F^2_{\times}(t) \right)
\end{equation}

We use the uncorrelated $T$ TDI channel (see Eq.~\ref{eq:COV}) to compute the LISA noise~\cite{Smith}. The LISA noise spectrum is characterized by two parameters: acceleration noise, $N_{acc} = 1.44 \times 10^{-48} \ \text{s}^{-4} \text{Hz}^{-1}$; and optical metrology system noise, $N_{pos} = 3.6 \times 10^{-41} \ \text{Hz}^{-1}$)~\cite{LISA}. We use a Fisher Information calculation $F_{ab}$ (see Eq.~\ref{eq:Fisher}) to estimate the uncertainty $\sqrt{F_{aa}^{-1}} = \sigma_a$ of the spectral separation (line plots in Fig.~\ref{fig:Uncertaintypro} are given for different observation durations), and independently an Adaptive-MCMC  (black scatter in Fig.~\ref{fig:Uncertaintypro}). This is a Metropolis-Hastings MCMC with a the following target proposal ($\Sigma_n$ the current empirical estimate of the covariance matrix, $\beta = 0.25$, $d$ the number of parameters, $N$ the multi-normal distribution). 
\begin{equation}
    Q_n(x)= (1-\beta)N(x,(2.28)^2 \Sigma_n / d ) + \beta N(x,(0.1)^2 I_d/d)
\end{equation}
\begin{equation}\label{eq:COV}
    \mathcal{C}(\theta,f) =
     \left(
     \begin{array}{ccc}
      S_A + N_A & 0 & 0  \\
      0 & S_E + N_E & 0 \\
      0 & 0 & N_T \\
     \end{array}
     \right)
   \end{equation}
 with $N_I(N_{acc}, N_{pos})$ \cite{Boileau}  the LISA noise of the channel $I = [A,E,T]$ and $S_I(f) = \frac{3H_0^2}{4 \pi^2} \frac{ \sum_i \Omega_{GW,i}}{f^3}$ the SGWB.
 
 The likelihood function ($\textbf{d}$ = data, $\theta$ = parameter) is given by Eq.~\ref{eq:likelihood}, and is used to estimate the posterior distribution $p(\theta|\textbf{d}) \propto p(\theta) \mathcal{L}(\textbf{d}|\theta)$; we use log uniform priors for magnitude parameters, and uniform priors  $p(\theta) = \prod_i U(\theta_i, a_i, b_i)$ to estimate SGWBs and LISA noise parameters.
\begin{equation} \label{eq:likelihood}
     \mathcal{L}(\textbf{d}|\theta) =  -\frac{1}{2} \sum_{k=0}^N \Bigg[ \frac{d_A^2}{S_A+N_A}  + \frac{d_E^2}{S_E+N_E} + \frac{d_T^2}{N_T} +  \ln\left(8\pi^3 (S_A+NA)(S_E+N_E)N_T \right) \Bigg] 
\end{equation}

\begin{equation} \label{eq:Fisher}
        F_{ab} = \frac{1}{2} \mathrm{Tr}\left(\mathcal{C}^{-1}\frac{\partial \mathcal{C}} {\partial \theta_a} \mathcal{C}^{-1} \frac{\partial \mathcal{C}}{\partial\theta_b}  \right) = \frac{1}{2} \sum_{I=A,E,T} \sum_{k=0}^N \frac{\frac{\partial S_I(f)+ N_I(f)} {\partial \theta_a}\frac{\partial S_I(f)+ N_I(f)} {\partial \theta_b}}{\left(S_I(f)+ N_I(f)\right)^2} 
\end{equation}

\section{Result}

\begin{figure}
\includegraphics[width=150mm]{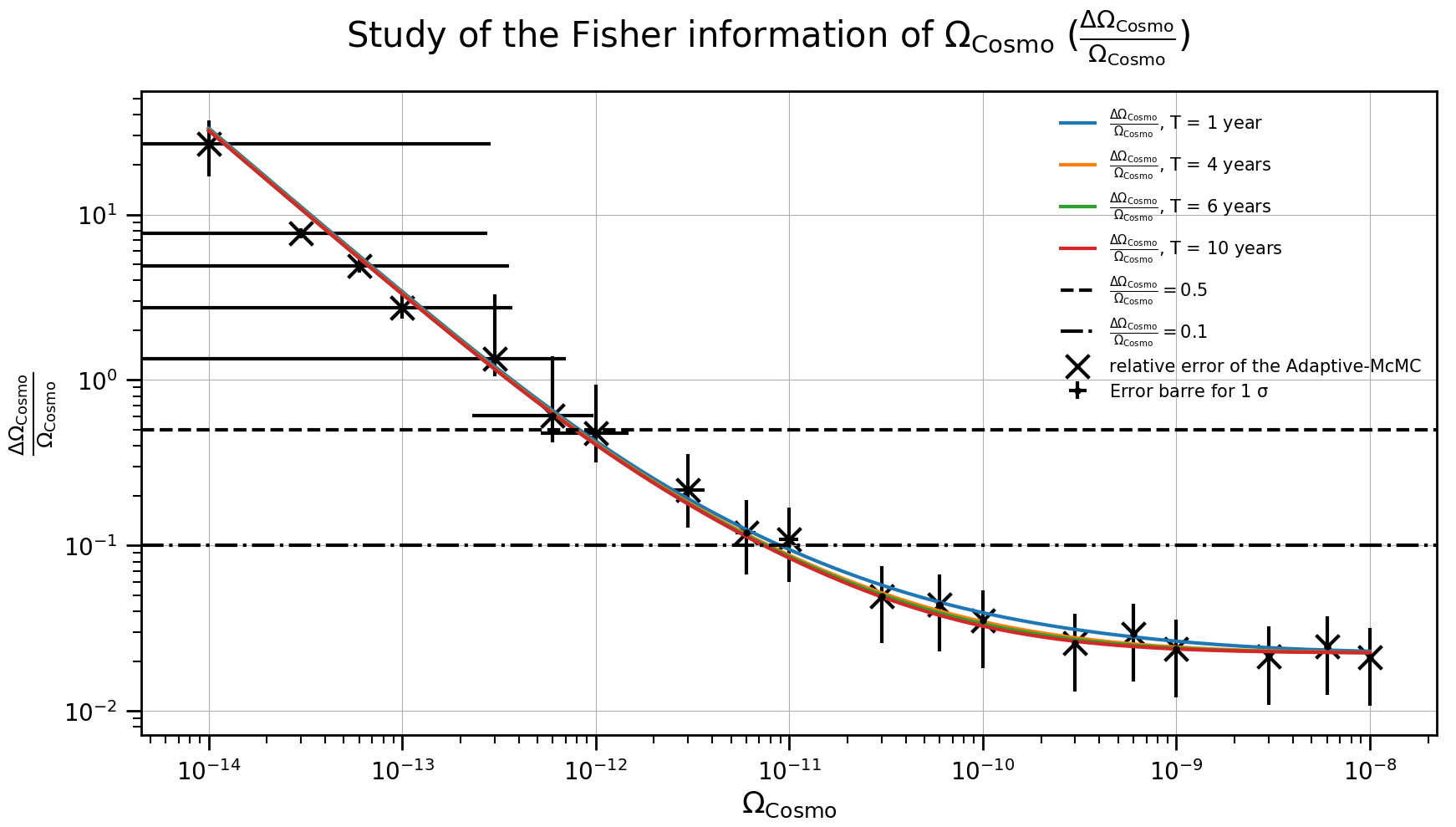}
\caption{Cosmological amplitude uncertainty estimates from the Fisher Iinformation study denoted by solid lines and from the MCMC by crosses. The upper horizontal dash line represent the error level $50\%$. In fact, above the line, the error is greater than $50\%$.}
\label{fig:Uncertaintypro}
\end{figure}

In Fig.~\ref{fig:Uncertaintypro} we display he uncertainty of the parameter estimates of $\Omega_0$ (the spectral energy density of the cosmological SGWB) from the Fisher Information study $F_{ab}^{-1} = \sigma_{ab}^2$ (displayed with lines), and from the A-MCMC (with the scatter points) for the channel $A$ and $E$, with the noise channel $T$ for 10 parameters $\theta = [N_{acc},$ $N_{pos},$ $\Omega_{astro},$ $\alpha_{astro},$ $\Omega_{cosmo},$ $\alpha_{cosmo},$ $A_1,$ $\alpha_1,$ $A_2,$ $\alpha_2 ]$. We conduct the study with different SGWBs,  astrophysical $\Omega_{GW,astro}$ and galactic $\Omega_{GW,galactic}$. The error bar is given for 1 standard deviation of a Gaussian posterior distribution.  The horizontal dashed line represents the error level of $50\%$; above this limit we cannot separate the cosmological SGWB. We use the LISA Model from Smith and Caldwell \cite{Smith} and the LISA parameters values from the proposal~\cite{LISA}.

\section{Conclusion}
We have studied the prediction of the measurement limit of a cosmological SGWB by LISA in the presence of an isotropic astrophysical background, a galactic foreground, and LISA noise for four years of data. The detection limit for DWD + BBH/BNS + Cosmo + LISA noise is $\Omega_{Cosmo,lim} = 8 \times 10^{-13}$~\cite{Boileau:2021sni}. In future work, we will study the spectral separation for more complex LISA noise, other DWD populations, and other cosmological models (broken power law, peaks in frequency).

\end{document}